# Ultrasensitive micro-scale parity-time-symmetric ring laser gyroscope

J. R EN ,[1] H. H ODAEI ,[1] G. H ARARI ,[2] A. U. H ASSAN ,[1] W. C HOW ,[3] M. S OLTANI ,[4] D. C HRISTODOULIDES ,[1] M. K HAJAVIKHAN[1,*]

[1]CREOL, The College of Optics and Photonics, University of Central Florida, Orlando, Florida 32816-2700, USA
[2]Physics Department and Solid State Institute, Technion, 32000 Haifa, Israel
[3] Sandia National Laboratories, Albuquerque, NM 87185-1086, USA
[4]Raytheon BBN Technologies, Cambridge, MA 02138, USA
*Corresponding author: mercedeh@creol.ucf.edu

**We propose a new scheme for ultrasensitive laser gyroscopes that utilizes the physics of exceptional points. By exploiting the properties of such non-Hermitian degeneracies, we show that the rotation-induced frequency splitting becomes proportional to the square root of the gyration speed ($\sqrt{\Omega}$)- thus enhancing the sensitivity to low angular rotations by orders of magnitudes. In addition, at its maximum sensitivity limit, the measurable spectral splitting is independent of the radius of the rings involved. Our work paves the way towards a new class of ultrasensitive miniature ring laser gyroscopes on chip.** © 2017 Optical Society of America

http://dx.doi.org/10.1364/OL.99.099999

In 1913, Sagnac demonstrated how the rate of rotation associated with an inertial frame of reference can be determined by optical means. In his experiments, the rotation speed was measured through the phase difference between two beams traveling in opposite directions within a loop. Since then, this approach has been successfully used to develop various families of optical rotational sensors [1,2]. A breakthrough in this area came shortly after the discovery of the laser, when Macek and Davis introduced gain in the ring cavity [3]. In this respect, the phase shift between the two counter-propagating beams is effectively converted into a splitting in the resonant frequencies that can in turn be readily measured.

In an ideal non-rotating ring laser, the two counter-propagating modes are expected to exhibit the same frequency. On the other hand, if this same system rotates at an angular frequency $\Omega$, the two initially degenerate resonant frequencies split, according to the following expression

$$\Delta\omega = \frac{8\pi A}{L\lambda}\Omega. \qquad (1)$$

Here $A$ and $L$ are the enclosed area and the perimeter of the ring, respectively, and $\lambda$ is the wavelength within the material associated with this cavity. Ideally, as long as the frequency separation ($\Delta\omega$) exceeds the quantum limit imposed by the spontaneous emission noise, the rotation speed $\Omega$ can be uniquely determined through a heterodyne measurement. For example, for a ring laser with a radius of 10 cm, operating at a wavelength of 1.55 μm, and rotating at a rate of $\sim 10°$/hour, one can expect a frequency splitting that is at best on the order of $\sim 12$ Hz [1]. In many consumer and industrial applications, it is required to detect angular velocities in the range of $\sim 0.1 - 100°$/hour - a precision that can be readily attained in state-of-the-art free-space ring laser gyroscopes [4]. Unfortunately however, such sensitivity levels have so far remained practically out-of-reach in fully integrated optical platforms, where the area of the loop is generally smaller, perhaps by several orders of magnitudes. In addition, in on-chip settings, light scattering from the cavity walls can prove detrimental. This is because the ensuing unwanted coupling between the two counter-propagating modes can lead to a lock-in effect, rendering this method ineffective below a certain rotation speed.

Currently, several efforts are underway to implement chip-scale laser gyroscopes based on different strategies. One possible approach is to detect rotation through the far-field emission of micro-scale deformed cavities [5]. Another technique makes use of narrow stimulated Brillouin Stokes lines in optically pumped silicon millimeter-scale disk resonators. Using this latter scheme, rotation measurements down to 15 °/hour have been demonstrated [6]. Yet another study suggests the use of gain threshold difference between the two counter-propagation modes in circular Bragg micro-lasers [7]. Optoelectronic semiconductor gain systems have also been considered for implementing electrically pumped ring laser gyroscopes. In such active III-V systems, the best detection rates reported so far were on the order of $\sim 100$ revolutions/sec ($\sim 10^8$ °/hour) [8]- although it has been theoretically predicted that centimeter-scale InP ring laser gyroscopes can detect rotations $\sim 180$ °/hour [9]. It has also been suggested that using a dual-cavity ring laser gyroscope could improve the detection limit, a possibility that is yet to be demonstrated in practice [10]. Ultimately, however, in such arrangements, the minimum detectable rotation rate is set by the size of the active ring- which is not easily scalable because of the large non-uniformities across most III-V wafers. In addition, semiconductor gain systems suffer from higher quantum noise levels due to carrier-induced index fluctuations [1]. In view of these issues, one may conclude that, without a significant enhancement of intrinsic sensitivity, the prospect of implementing on-chip RLGs for detecting rotation rates on the order of $\Omega \sim 0.1 - 100\ °$/hour is very challenging.

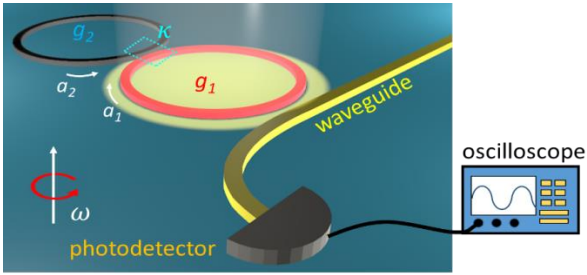

Fig. 1. Schematic of a parity-time-symmetric laser gyroscope system. The light generated in the resonators is coupled to waveguide that carries it to a fast photodiode. The beat frequency can be used to determine the rotation rate.

Here we propose a novel class of ring laser gyroscopes based on the physics of non-Hermitian degeneracies. By exploiting the properties of exceptional points in judiciously designed parity-time-symmetric-type arrangements, we show that the ensuing frequency splitting is proportional to the square-root of the gyration speed ($\sqrt{\Omega}$)- an effect that can boost the sensitivity to small rotations by orders of magnitudes [11-17]. Moreover, when the maximum sensitivity limit is reached, the splitting is no longer proportional to the radii of the rings involved. Using our technique, the lock-in effect can be entirely avoided by enforcing directional propagation in each ring. We will also show that the strong bifurcation around a non-Hermitian degeneracy can be utilized to actively reposition this system back to its optimum operating point for rotation measurement.

Figure 1 depicts a schematic of the proposed non-Hermitian gyroscope. At the heart of this system is a pair of coupled ring resonators (of radius R and coupling strength κ). The rings are identical in every respect but are subject to different levels of gain and/or loss. The contrast between the amplification levels experienced by the rings can be introduced for example through preferential pumping. Although this system is not strictly invariant under the simultaneous action of parity (P) and time (T) operators, it becomes PT symmetric once it is gauged by a constant gain/loss bias [18-25]. In a rotating frame, one may assume that the beam propagating in a clockwise (CW) direction in one of the rings acquires an effective Sagnac phase shift with respect to a counter clockwise (CCW) wave of the same frequency circulating in the other. A bus waveguide may be placed on the side of one (or both) ring(s) to direct the lasing emission into a photodiode in order to measure the resulting beat frequency.

The mechanism behind the enhanced sensitivity in the proposed PT symmetric coupled cavity configuration can be explained by considering the modal behavior of the system [19]. In general, each ring, when uncoupled, can support a number of longitudinal modes both in the clockwise and counterclockwise directions. Without loss of generality, here we limit our analysis to a single longitudinal mode in one direction [25, 26]. The cross-section of the rings can be designed so as to support only the fundamental TE mode. Unidirectional light propagation can be enforced in the rings through a geometrical design, for example using an s-bend bypass in the rings as previously demonstrated in [26]. In order to analyze this system, we consider the interaction between the clockwise field in one cavity and the counterclockwise traveling wave in the neighboring resonator. In this respect, the interplay between the electric modal fields in the two identical rings can be effectively described through a set of time dependent coupled equations:

$$\begin{cases} i\dot{a}_1 + \omega_1 a_1 - ig_1 a_1 + \kappa a_2 = 0 \\ i\dot{a}_2 + \omega_2 a_2 - ig_2 a_2 + \kappa a_1 = 0 \end{cases} \quad (2)$$

where $a_1, a_2$ represent the modal amplitudes in the two cavities. The angular frequencies, $\omega_1$ and $\omega_2$, are independently determined by the resonance conditions for each resonator in the absence of coupling. For two identical cavities at rest, these frequencies are expected to be the same ($\omega_1 = \omega_2 = \omega_0$). The gain (loss) in each ring is denoted by $g_1$ and $g_2$, respectively. One can show that the eigenmodes of this system and their corresponding eigenfrequencies are given by:

$$|1\rangle = [1\ \ e^{i\theta}]^T \qquad |2\rangle = [1\ \ -e^{-i\theta}]^T$$
$$\omega_{PT_{1,2}} = i\omega_0 + \left(\frac{g_1+g_2}{2}\right) \pm i\sqrt{\kappa^2 - \left(\frac{g_1-g_2}{2}\right)^2} \quad (3)$$

where $\theta = \sin^{-1}((g_1 - g_2)/2\kappa)$. While for a certain range of values of $\kappa$, this system will support two distinct modes, Eq. (3) shows that at the vicinity of $2\kappa = |g_1 - g_2|$, the dimensionality of the system abruptly collapses. At this point, not only the two eigenfrequencies coalesce, but also the eigenvectors become identical. As a result, the system lases in only one frequency $\omega_0$. This type of degeneracy (that is unique to non-Hermitian arrangements) marks the onset of a phase transition. The location in parameter space in which this degeneracy occurs is known as an exceptional point. The solid lines in Fig. 2 show the trajectory of the eigenvalues in the complex plane. As the ratio ($|(g_1 - g_2)/\kappa|$) is varied, the system undergoes a dramatic transition. Note, that a global change in the gain (loss) in this two-ring configuration will only deform the trajectories, yet but the overall trend remains unchanged.

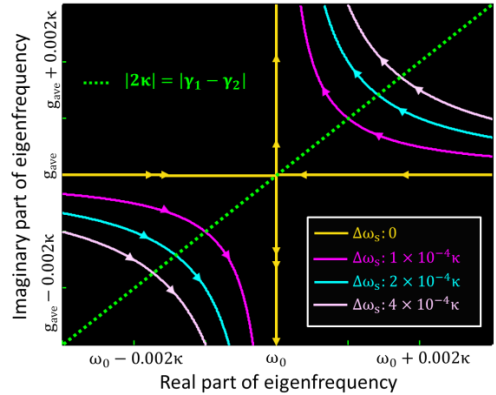

Fig. 2. The trajectories of the eigenfrequencies ($i\omega_{PT}$) in the complex plane for different detuning levels ($\Delta\omega_s$). The dotted line represent the condition where $|2\kappa| = |g_1 - g_2|$. For the perturbed systems, the distance between the intersections of the associated curves (depicted by the same color) and the dotted line represents the splitting between the corresponding eigenfrequencies. The exceptional point is located exactly at the center of the plane.

Let us now consider this same system in a rotating frame having an angular frequency $\Omega$. If the magnitude of the maximum speed ($v_{max} = \Omega R$) is small compared to the speed of light, one can then ignore relativistic effects. Under these conditions, in a single ring, the corresponding resonant frequency is expected to change by an amount $\Delta\omega_s$ due to the Sagnac shift:

$$\Delta\omega_s = \frac{\omega_0 n R \Omega}{c} . \quad (4)$$

In other words, in this rotating frame, the two resonant frequencies associated with these rings (when uncoupled), will be modified according to $\omega_1 = \omega_0 + \Delta\omega_s$ and $\omega_2 = \omega_0 - \Delta\omega_s$. For a system that was originally placed at the exceptional point ($2\kappa = |g_1 - g_2|$), the presence of such rotation-induced detuning perturbations, will give rise to a new set of eigen-solutions:

$$|1\rangle = \begin{bmatrix} 1 & e^{i\theta'} \end{bmatrix}^T \quad |2\rangle = \begin{bmatrix} 1 & -e^{-i\theta'} \end{bmatrix}^T$$

$$\omega_{PT_{1,2}\,perturbed} = i\omega_0 + \left(\frac{g_1+g_2}{2}\right) \pm i\sqrt{\kappa^2 - \left(\frac{g_1-g_2+2i\Delta\omega_s}{2}\right)^2} \quad (5)$$

where $\theta' = \sin^{-1}((g_1 - g_2 + 2i\Delta\omega_s)/2\kappa)$. Equation (5) clearly demonstrates that the rotation induced detuning forces the system to depart from the exceptional point. This implies that, even at $2\kappa = |g_1 - g_2|$, once it is perturbed, this arrangement will now support two supermodes with a beat frequency $\Delta\omega_{PT} = |\Im\{\omega_{PT_1} - \omega_{PT_2}\}|$. For small rotation velocities where $|\Delta\omega_s| \ll |g_1 - g_2|$, these two eigenvalues are approximately given by:

$$\omega_{PT_{1,2}\,perturbed} \cong i\omega_0 + \left(\frac{g_1+g_2}{2}\right) \mp (1-i)\sqrt{\Delta\omega_s \kappa} \quad (6)$$

Consequently, the splitting between the real components of these two eigenfrequencies ($i\omega_{PT_{1,2}}$) is now expressed by:

$$\Delta\omega_{PT\,perturbed} \cong 2\sqrt{|\Delta\omega_s \kappa|} \quad (7).$$

Equation (7) confirms that the beat frequency in the parity-time-symmetric ring gyroscope has a square-root-dependence on $\Delta\omega_s$. For small rotation rates ($\Delta\omega_s \ll 1$), this square-root behavior can indeed result in a substantially increased frequency separation. This behavior is also evident in Fig. 2 where the trajectories of the eigenfrequencies are depicted for several detuning levels. The green dotted line represents the locus of points at which $2\kappa = |g_1 - g_2|$ is satisfied. The distance between the intersections of the iso-color curves and the green dotted line indicates the amount of splitting between the corresponding complex eigenfrequencies.

It should be noted that the square-root-dependence of the beat frequency on externally induced perturbations is universal in all non-conservative systems that are operating around an exceptional point [12,16]. However, in coupled cavity parity-time-symmetric-like configurations, this response is further scaled by the square root of the coupling strength $\kappa$ [18]. For a pair of identical resonators, under the weak-coupling approximation, the coupling factor can in principle be as large as a quarter of the free spectral range i.e. $\kappa_{max} = c/8\pi R n_g$, here $n_g$ is the group index. By inserting this value in Eq. (7), the maximum attainable beat frequency in this arrangement is:

$$\Delta\omega_{PT\,max} = \sqrt{\frac{n}{n_g}\frac{\omega_0}{2\pi}\Omega} \quad (8).$$

Remarkably, Eq. (8) shows that unlike standard ring laser gyroscopes, the maximum frequency splitting is now completely independent of the radius of the rings involved. In this respect, one can now envision a micro-scale ring laser gyroscope that in principle can exhibit a sensitivity similar to that obtained in centimeter-long systems. Of course, for smaller rings, $\kappa_{max}$ is larger; this has to be compensated by utilizing higher levels of gain-loss contrast- something that is readily available in semiconductor active media.

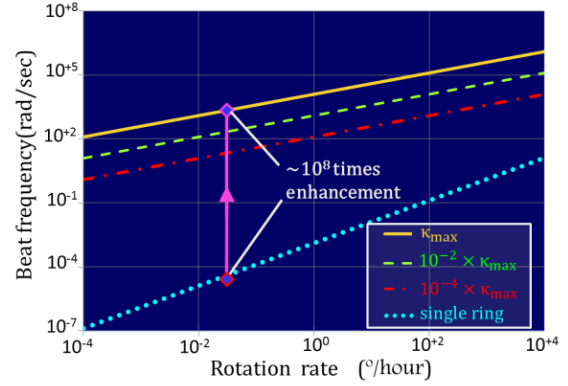

Fig. 3. The beat frequency as a function of the rotation rate for a single ring (dotted line) and the PT symmetric coupled ring systems with various coupling strength levels.

It is instructive to compare the rotation-induced frequency splitting in a PT-system ($\Delta\omega_{PT}$) and the beat frequency associated with a pair of counter-propagating modes in a single ring ($\Delta\omega = 2\Delta\omega_s$). In particular, one can define a sensitivity enhancement (S.E.) factor as the ratio $\Delta\omega_{PT}/\Delta\omega$:

$$S.E. = \sqrt{\left|\frac{2\kappa}{\Delta\omega}\right|} \quad (9)$$

This coefficient may be viewed as the *scale factor* for the PT-symmetric ring laser gyroscope. The sensitivity enhancement (*S.E.*) depends on the coupling strength and the rotation rate ($\Delta\Omega$). As an example, if the coupling strength is $\kappa \sim 10^{12}\,s^{-1}$, and if $\Delta\omega=1$ Hz, then one can expect a $S.E. \cong 10^6$. This implies that a rotation of $\Omega = 100\,°/hour$ that could at best generate a frequency splitting on the order of $\sim 0.6$ Hz in a single ring with a radius of a 100 $\mu m$, now in a two-ring PT system, can result into a splitting of $\sim 1.8$ MHz- an improvement of more than six orders of magnitude. To further elucidate these aspects, Fig. 3 compares the performance of a two-ring PT symmetric arrangement to that of a single ring. The radii of the rings are here taken to be $R = 100\,\mu m$, and the operating wavelength is centered at $\lambda_0 = 1.55\,\mu m$. Figure 3 shows the beat frequency as a function of the rotation rate in a log-log scale. The dotted line with a slope of unity represents the beat frequency expected from a conventional ring laser gyroscope. On the other hand, in a non-Hermitian PT-symmetric system, the slope of the line drops to one half - indicating a superior square-root behavior.

While most free-space ring laser gyroscopes are based on counter propagating modes in a single ring, for on-chip laser gyroscopes, a double ring (or race-track) configuration has been speculated as a favorable geometry [10]. This is mainly because in waveguide-based lasers, unlike existing free-space ring laser gyroscopes, the scattering from the walls is more pronounced and less preventable. The scattering couples the two counter propagating modes and makes it difficult to deduce the rotation rates below the lock-in limit. In this respect, an important advantage of a double ring arrangement is that the beating mechanism is no longer between the two counter-propagating modes of the same ring. In a coupled ring geometry, the modes in the two rings can be contrasted with respect to each other. As a result, one can avoid the complications arising due to lock-in effects by designing the rings in such a way that they inherently support modes in a unidirectional fashion. For on-chip single ring lasers, there are currently a number of techniques to suppress one of the counter-propagating modes at the expense of the other. One example is to use an *s*-bend cavity structure as first suggested by Hohimer [26]. For larger rotation rates, where lock-in effects are not a concern, the rings may be designed to support two counter-propagating modes. In addition, to improve the detection limit, one may also want to consider designs that permit high coherence emission like that reported in [27].

In designing sensors with large scale-factors, one important consideration is the sensor's response to unwanted drifts of the parameters, either due to environmental variations or intrinsic noise effects. Clearly, the proposed gyroscope, involving coupling, gain-contrast, and possibly some detuning between its constituent elements, is expected to react in a complex fashion to such deviations from its nominal parameters. A preliminary error analysis indicates that this device can reach its full potential for sensing purposes only if it operates at or close to an exceptional point. In fact, if it is biased far away from this point ($\kappa^2 - (\Delta g/2)^2 \neq 0$), the uncertainty ($\delta\Omega$) in rotation rate ($\Omega$) will be directly proportional to the coupling $\kappa$ ($\delta\Omega/\Omega = 2\kappa \, \delta\kappa/\sqrt{\Delta\omega_s}$). Since $\kappa$ is typically designed to have large values ($\sim 10^{12} \, sec^{-1}$), operating in this regime could lead to very large errors, rendering this technique ineffective. It is, therefore, crucial to constantly reposition the system at the exceptional point, despite of the ever-changing environmental factors. Fortunately, exceptional points are well-defined features in parameter space- since at these junctures the arrangement undergoes an abrupt phase transition. Consequently, such points can be readily identified by monitoring the variation of the observable ($\delta\Delta\omega_{PT}$) with respect to a scanning parameter ($\Delta g$ or $\kappa$). Whether the arrangement is at rest or already experiencing rotation, at the vicinity of this point, the absolute change in the $\Delta\omega_{PT}$ reaches a maximum. One can then choose the measurement result performed at this extremum point to be the most accurate value for the rotation rate. It should be noted that the existence of such a well-defined reference point is crucial for our proposed device. In the past decade, there have been a number of proposals for *passive* on-chip gyroscopes based on fast-light effects [28] in coupled resonators- a property that can also lead to large scale-factors. However, due to the lack of such reference points, these systems are susceptible to the drift of their parameters. Once at the exceptional point, the uncertainties in coupling/gain-contrast can at most generate the same degree of error in the rotation rate ($\delta\Omega/\Omega = \delta\kappa/\kappa = \delta g/g$).

In conclusion, in this paper, we have introduced a new physical principle to increase the Sagnac effect and we show how it can be used to develop a new family of ring laser gyroscopes. The enhanced sensitivity in our system is attributed to the intriguing properties of exceptional points since at these junctures the system undergoes an abrupt change in eigen-space. At such a point, the system responds to external perturbations in a square-root fashion. For small rotation rates, this could result in orders of magnitude enhancement in sensitivity. We have also shown that when this effect is fully utilized, the frequency splitting becomes independent of the size of the rings involved- hence, our approach can be used to realize micro-scale ring laser gyroscopes with the same sensitivities as those at cm-scale. Finally, the system can be made resilient to drift of its parameters, since it can be reliably brought back to the exceptional point- where an accurate measurement can be performed. This approach is generally applicable in any system with gain- even if it is of parametric type. The proposed approach may open new directions towards the realization of highly sensitive, on-chip, and miniature ring laser gyroscopes.

**Funding.** Office of Naval Research (ONR) (N00014-16-1-2640); National Science Foundation (NSF) (ECCS-1128520, ECCS-1454531); Army Research Office (ARO)(W911NF-14-1-0543, W911NF-16-1-0013); Air Force Office of Scientific Research (AFOSR)(FA9550-14-1-0037).